\documentstyle[11pt,newpasp,twoside]{article}
\def\edcomment#1{\iffalse\marginpar{\raggedright\sl#1\/}\else\relax\fi}
\reversemarginpar

\begin{document}
\title{The JVAS/CLASS $6^{\prime\prime}$ to $15^{\prime\prime}$ Lens Search}
 \author{P.~M.~Phillips, I.~W.~A.~Browne, P.~N.~Wilkinson, N.~J.~Jackson}
\affil{University~of~Manchester, Jodrell~Bank~Observatory, Macclesfield, Cheshire, U.K. SK11~9DL}
\author{and members of the CLASS collaboration}

\begin{abstract}
We present the strategy and status of a gravitational lens search for multiple imaging with angular separations between $6^{\prime\prime}$ and $15^{\prime\prime}$ within the combined JVAS and CLASS dataset of $\sim15\,000$ flat-spectrum radio sources. Currently all but one lens candidate have been rejected.
\end{abstract}

\section{Introduction}
The Jodrell-Bank VLA Astrometric Survey (JVAS) and the Cosmic Lens All Sky Survey (CLASS) samples together comprise $\sim15\,000$ radio sources in the northern sky selected to be flat-spectrum between 1.4~GHz and 5~GHz (see the contribution by Browne \& Myers). The simple compact structures of such sources ease the task of identifying potential gravitational lensing events. So far, a search for $0.\!\!^{\prime\prime}3$ to $6^{\prime\prime}$ scale lensing (produced by galaxy-scale masses) has been particularly fruitful, with at least 18 gravitational lens systems found within the two samples. This poster describes a search for further lensing events on larger angular scales, corresponding to lensing by  groups and clusters of galaxies.

\section{Why and How?}
Statistics on the prevalence of galaxy lensing events relate to the clumpiness of matter in the universe and provide constraints on the cosmological parameters. The frequency of lensing by galaxy-sized masses is dominated by the cosmological constant rather than the density parameter, because the number density of lensing masses is assumed to be constant. Searching for lensing by groups and clusters provides a different perspective on these parameters because the relevant mass scales are evolving at redshifts where lensing is most probable. 

A systematic search for lensing has the advantage of tracing not only visible matter, but also dark matter. Hawkins (1997) suggested that there is a significant proportion of dark galaxies producing image splittings as large as $8^{\prime\prime}$. However, Jackson et al. (1998) have detected all the lensing galaxies within a sample of 12 lenses from JVAS and CLASS, although the image separation upper limit of $6^{\prime\prime}$ would have selected against {\bf massive} dark galaxies. Kochanek et al. (1999) also argue against Hawkins based on the number of radio and optical pairs seen, and suggest that these `large' separation pairs are binary quasars. The result of an extended search of JVAS and CLASS will provide a clean sample to test these hypotheses.

The JVAS/CLASS data consist of VLA A-configuration 8.4~GHz snapshot observations. The data were systematically calibrated and automapped, and candidate lens systems were chosen with component separations between $6^{\prime\prime}$ and $15^{\prime\prime}$ using the same methodology as the $0.\!\!^{\prime\prime}3$ to $6^{\prime\prime}$ lens search; i.e. only systems with multiple compact components were selected. A total of 10 systems met these initial selection criteria.

The VLA was again used in A-configuration, this time at 1.4~GHz and 15~GHz in order to reveal any low surface brightness emission which could allow the identification of intrinsic double sources (i.e. AGNs), and to provide higher resolution images. The combined flux density information from these two frequencies and the original 8.4~GHz observations provided spectral information for each component which could also be used to reject a system as a lens candidate. A total of 3 candidates remained after this stage.

Higher resolution radio maps of the three remaining candidates were obtained with MERLIN at 1.6~GHz. One candidate could be rejected based on morphological grounds, whilst another was found to have a significantly different percentage polarisation and surface brightness between each component and so could also be rejected. Only one candidate remains after this stage.

\section{Current Work}
We have recently obtained WHT service time observations of the remaining candidate. A preliminary analysis suggests that each component is detected, although the optical component flux density ratio is markedly different from the radio, and the weaker component appears to be slightly extended. If it is shown not to be a lens system, the fact that no $6^{\prime\prime}$ to $15^{\prime\prime}$ scale lensing events are found from $\sim15\,000$ background sources may rule out the Standard Cold Dark Matter model, however more realistic simulations of mass distributions on these scales must be made before significant constraints can be placed on the cosmological parameters.

\acknowledgements
The VLA is the Very Large Array and is operated by the National Radio Astronomy Observatory which is a facility of the National Science Foundation operated under cooperative agreement by Associated Universities, Inc. MERLIN is the Multi-Element Radio Linked Interferometer Network and is a national facility operated by the University of Manchester on behalf of PPARC. This research was supported in part by European Commission, TMR Programme, Research Network Contract ERBFMRXCT96-0034 ``CERES''. P.~M.~Phillips would also like to thank PPARC for the support of a studentship award.

\end{document}